\documentclass[leqno,a4paper,12pt,oneside]{article}
\usepackage[latin1]{inputenc}
\usepackage[T1]{fontenc}

\usepackage[italian]{babel}
\usepackage[mathscr]{euscript}
\usepackage{amsmath,exscale,latexsym,amssymb,amsthm,amsfonts}

\title{A possible quantic motivation of the structure of quantum group}
\author{Giuseppe Iurato}
\date{}
\begin{document}
\maketitle
\begin{quotation}\small\bf Abstract. \it Following a suggestion of A. Connes \rm(\it see \rm [Co], § I.1),
\it we build up a (first) simple natural structure of a no finitely generated braided non-commutative Hopf
Algebra, suggested by elementary quantum mechanics.\end{quotation}\normalsize

\subsection*{1. The notion of EBB-groupoid}The idea of \it quantum group \rm has been introduced, in a pure
mathematical context, independently by V.G. Drinfeld\footnote{Drinfeld put, as basic examples of a quantum group
structure, some group algebras dually related with certain function algebras, starting from simple quantum
mechanics considerations (see [Dr2], § 1).} ([Dr1], [Dr2]) and M. Jimbo ([Ji1], [Ji2]), who used the adjective
'quantum' for the fact that a such structure is obtained quantizing (according to geometric quantization) a
Poisson symplectic structure introduced on the algebra $\mathfrak{F}(G,\mathbb{C})$ of the differentiable
$\mathbb{C}$-functions defined on a Lie group $G$. This structure of quantum group is the result of a non-
commutative Hopf algebra achieved as non-commutative (non-trivial) deformations ([BFFLS]) of the initial Hopf
algebra $\mathfrak{F}(G,\mathbb{C})$, or of the universal enveloping Hopf algebras of an arbitrary semisimple
Lie algebras. A little later, Yu. Manin (see [Mn]) and S.L. Woronowicz ([Wo]) independently constructed
non-commutative deformations of the algebra of functions on $SL_2(\mathbb{C})$ and $SU_2$, respectively.\\On the
other hand, the study of quantum group was inspired by the works of physicists on integrable XYZ models with
highest spin. Moreover, L.D. Faddeev and coworkers ([Fa], [FST], [FT]), had already used similar structures in
mathematical physics, for the algebra of quantum inverse scattering transform of the theory of integrable
models, {as well as} P.P. Kulish, N.Yu. Reshetikhin ([KR]) and E.K. Sklyanin ([KuS]).\\\\However, from
elementary quantum mechanics, it is possible constructs a simple model of a braided non-commutative Hopf
algebra, through an algebraically modern re-examination of few elementary notions of the original papers of W.
Heisenberg, M. Born e P. Jordan ([33]) on matrix quantum mechanics.\\\\The basic algebraic structure for our
model is that of \it groupoid. \rm Independently, A. Baer ([Ba]), and W. Brandt ([Br]) in some of his researches
on quadratic forms on $\mathbb{Z}$, introduced a well-defined new algebraic structure, that we will call \it
Ehresmann-Baer-Brandt groupoid \rm(or \it EBB-groupoid\rm) (see [Co3], [CW], [La], [Hi], [Mak]). It is a partial
algebraic structure because is an algebraic system provided by a partial (or no total) composition law, that is,
not defined everywhere.\\\\Historically, S. Eilenberg and N. Steenrod ([ES]) formulated the notion of abstract
category on the basis of the previous notion introduced by S. Eilenberg and S. MacLane ([EM]), considering a
groupoid as a special category in which every morphism has an inverse (see also [MacL]).\\A little later, the
notions of abstract category and groupoid appeared too in the work of C. Ehresmann ([Eh]).\\However,
retrospectively, the common substructure to all these notions of groupoid, is that of \it multiplicative system
\rm(or \it Eilenberg-Steenrod groupoid, \rm or \it ES-groupoid\rm), say $(G,\mathcal{D}_{\diamond},\diamond)$,
\rm where $\diamond$ is a partial composition law over $G$, defined on a domain $\mathcal{D}_{\diamond}\subseteq
G\times G$. It was introduced by [ES].\\ The further categorial (over) structure of a ES-groupoid, leads to the
notion of \it Eilenberg-MacLane groupoid \rm(or \it EM-groupoid\rm), introduced in [EM], and to the
equivalent\footnote{See [Wa] for a proof of this equivalence.} notion of \it Ehresmann groupoid \rm (or \it
E-groupoid\rm), introduced in [Eh]. We will use the notion of E-groupoid.\\\\An \it E-groupoid \rm is an
algebraic system of the type $(G,G^{(0)},r,s,\star)$, where $G,G^{(0)}$ are non-void sets such that
$G^{(0)}\subseteq G$, $r,s:G\rightarrow G^{(0)}$ and $\star:G^{(2)}\rightarrow G$ with $G^{(2)}=\{(g_1,g_2)\in
G\times G, s(g_1)=r(g_2)\}$, satisfying the following conditions:\begin{itemize}\item$\!\!\!_1$ $s(g_1\star
g_2)=s(g_2)$, $r(g_1\star g_2)=r(g_1)$\ \ $\forall (g_1,g_2)\in G^{(2)}$;\item$\!\!\!_2$ $s(g)=r(g)=g\ \ \forall
g\in G^{(0)}$;\item$\!\!\!_3$ $g\star\alpha(s(g))=\alpha(r(g))\star g=g\ \ \forall g\in G$;\item$\!\!\!_4$
$(g_1\star g_2)\star g_3=g_1\star(g_2\star g_3)\ \ \forall g_1,g_2,g_3\in G$;\item$\!\!\!_5$ $\forall g\in G,\
\exists g^{-1}\in G:\ g\star g^{-1}=\alpha(r(g)),\ g^{-1}\star g=\alpha(s(g))$,\end{itemize}where
$\alpha:G^{(0)}\hookrightarrow G$ is the immersion of $G^{(0)}$ into $G$. The maps $r,s$ are called,
respectively, \it range \rm(or \it target\rm) and \it source\rm, $G$ the \it support \rm and $G^{(0)}$ the \it
set of units \rm of the groupoid. $g^{-1}$ is said to be the \rm(bilateral) \it {inverse} \rm of $g$, so that we
have an inversion map of the type $i_G:g\rightarrow g^{-1}$, defined on the whole of $G$. Instead, the map
$\star$ is a partial map defined on $G^{(2)}\subseteq G\times G$, and not on the whole of $G\times G$. However,
from now on, for simplicity, we will suppresses the symbol $\alpha$ in 3. and 5., writing only $r(g),s(g)$
instead of $\alpha(r(g)),\alpha(s(g))$.\\However, a E-groupoid is more general than the first notion of groupoid
used by W. Brandt and A. Baer, that we call \it {Baer-Brandt groupoid} \rm(or \it {BB-groupoid}\rm); indeed, a
BB-groupoid \rm may be defined as a E-groupoid satisfying the further condition\begin{itemize}\item$\!\!\!_6$
$\forall g,g''\in G, \ \ \exists g'\in G\ \mbox{such\ that}\ (g,g')\in G^{(2)},\ (g',g'')\in
G^{(2)}$,\end{itemize}so that, we will call an \it {Ehresmann-Baer-Brandt groupoid}\rm (or \it
{EBB-groupoid}\footnote{A. Nijenhuis ([Ni]) sets the structures of groupoid in the general context of the group
theory, whereas a topological characterizations of such structures is due to S. Golab ([Go]).}\rm), an algebraic
system satisfying $\bullet_i\ \ i=1,...,6$.\\\\In view of the applications to quantum mechanics, it is important
the following particular examples of EBB-groupoid (see [Co3], § II.5).\\If $X$ is an abstract (non-empty) set,
let $G=X^2$, $G^{(0)}=\Delta(X^2)= \{(x,x);\ x\in X\}$, $r=pr_1:(x,y)\rightarrow x,\ s=pr_2:(x,y)\rightarrow y\
\ \forall x,y\in X$ and $(x,y)\star(y,z)=(x,z)$. Then, it is easy verify that $(G,G^{(0)},r,s,\star)$ is a
EBB-groupoid, with $(x,y)^{-1}_d=(x,y)^{-1}_s=(y,x)$. It is called\footnote{Or \it {pair EBB-groupoid}, \rm or
\it {coarse EBB-groupoid}.} the \it natural EBB-groupoid \rm on $X$, and denoted by ${\cal G}_{EBB}(X)$.\\On the
other hand, there exists E-groupoids that are not EBB-groupoids. For instance, if $\Gamma$ is a group that acts
on a set $X$ by the action $\psi:\Gamma\times X\rightarrow X$, if $e$ is the identity of $\Gamma$, and if we
put\footnote{In the definition of $r$ and $s$, it is necessary to consider the bijection $x\rightarrow (e,x)\ \
\forall x\in X$, that identifies $X$ with $\{e\}\times X$.} $G=\Gamma\times X,\ G^{(0)}=\{e\}\times X,\
r(g,x)=x,\ s(g,x)=\psi(g,x)$ and $(g_1,x)\star(g_2,y)=(g_1g_2,x)$ if and only if\footnote{Indeed,
$G^{(2)}=\{((g_1,x),(g_2,y))\in G^2;\ s(g_1,x)=r(g_2,y)\}$ and $s(g_1,x)=\psi(g_1,x)=g_1x=y=r(g_2,y)$, that is
$y=g_1x$.} $y=\psi(g_1,x)$, then we obtain a E-groupoid, with $(g,x)^{-1}=(g^{-1},\psi(g,x))$, called the \it
semi-direct product E-groupoid \rm of $\Gamma$ by $X$ \rm and denoted by $\Gamma\ltimes_EX$.\\ Nevertheless,
such a structure cannot be a EBB-groupoid if, for example, the action $\psi$ is not transitive. In fact, for
every $(g_1,x),(g_3,z)\in G$, there exists $(g_2,y)\in G$ such that $(g_1,x)\star(g_2,y)\in G^{(2)}$ and
$(g_2,y)\star(g_3,z)\in G^{(2)}$ if and only if $y=\psi(g_1,x)$ and $z=\psi(g_2,y)$, that is, if and only if
there exists $g_2\in G$ such that $z=\psi(g_2,\psi(g_1,x))$ for given $z,\psi(g_1,x)\in X$. Therefore, if $\psi$
is a no transitive action, follows that $\Gamma\ltimes_EX$ is a E-groupoid but not a EBB-groupoid.\\\\The
following result puts in relation the two structures of E-groupoid and EBB-groupoid (see [Wa]).\\\\\bf I. \it A
multiplicative system $(G,\mathcal{D}_{\diamond},\diamond)$ is a E-groupoid if and only if there exists a unique
partition $\mathcal{P}$ of $G$ such that $\mathcal{D}_{\diamond}\subseteq\bigcup_{A\in\mathcal{P}}A$ and that
the induced multiplicative system $(A,\diamond_A)$ be a EBB-groupoid for every $A\in\mathcal{P}$.\\\\\rm For
instance, if the given multiplicative system is a E-groupoid, namely $(G,G^{(0)},r,s,\star)$ (with
$\star=\diamond$), then we consider the relation$$\sim\ \doteq\{(g,g'')\in G\times G;\ \exists g'\in G,\
(g,g')\in G^{(2)},\ (g',g'')\in G^{(2)}\}.$$Hence, it is possible to prove that $\sim$ is an equivalence
relation, so that $\mathcal{P}=G/\sim$ is a partition of $G$. If we set $A^{(2)}=G^{(2)}\cap A$,
$A^{(0)}=G^{(0)}\cap A$, $r_A=r_{|_A},s_A=s_{|A}$ and $\star_A=\star_{|_{A^{(2)}}}$, for every
$A\in\mathcal{P}$, then $(A,A^{(0)},r_A,s_A,\star_A)$ is a EBB-groupoid.\subsection*{2. The notion of
convolution structure\footnote{See [Gr], [No], [AM], [Vr].}}Let $A$ be an unitary commutative ring, $J$ a set of
indices and $X=\{x_j;\ j\in J\}=(x_j)_{j\in J}$ a family of abstract symbols. Let$$\langle
X\rangle=\langle(x_j)_{j\in J}\rangle=\{\sum_{j\in J}a_jx_j;\ a_j\in A,\ x_j\in X\}$$be the set of the formal
linear combinations of the elements of $(x_j)_{j\in J}$ with coefficients on $A$: $a_jx_j$ must be understood as
the value of a map of the type $A\times X\rightarrow\langle X\rangle$, whereas $\sum_{j\in J}a_jx_j$ have all
coefficients $a_j$ zero except a finite number; $\langle X\rangle$ is said to be the \it free set \rm generated
by $X$. With the operations$$\big(\sum_{j\in J}a_jx_j\big)+\big(\sum_{j\in J}b_jx_j\big)=\sum_{j\in
J}(a_j+b_j)x_j,\leqno(5)$$
$$a\cdot\big(\sum_{j\in J}a_jx_j\big)=\sum_{j\in J}(aa_j)x_j,\ \ a\in A,\leqno(6)$$ $\langle X\rangle$ is a $A$-module,
said to be the \it {free $A$-module} \rm on $X$, with \it {base} X. \rm In this context, it is possible to prove
that its elements admits a unique decompositions of the type $\sum_{j\in J}a_jx_j$. We will denote this free
$A$-module by ${\cal M}_A(X)=(\langle X\rangle,+,\cdot)$, the operations $+,\cdot$ being (5), (6). Furthermore,
it is possible to prove that: a $A$-module $(M,+,\cdot)$ is free on $(\emptyset\neq)X\subseteq M$ if and only
if, for any $A$-module $(M',+,\cdot)$ and for any map $\varphi:X\rightarrow M'$, there exists a unique
$A$-homomorphism of $A$-modules $\psi:M\rightarrow M'$ such that $\varphi=\psi{\big |}_{X}$ (in such
\underline{a} case, $\psi$ is said to be an \it $A$-extension \rm of $\varphi$, whereas $X$ is a base of $M$,
that is $M=\langle X\rangle$). A base identifies a unique free $A$-module in the sense that, if $M,M'$ are two
free $A$-modules with respective bases $X,X'$ and $f:X\rightarrow X'$ is a bijection, then $M,M'$ are
$A$-isomorphic.\\These last results are important in the processes of linear extension from a base.\\\\On the
free $A$-module ${\cal M}_A(X)=(\langle X\rangle,+,\cdot)$ it is possible to establish a structure of unitary
ring (in general, non-commutative) by a precise\footnote{A priori, the formal choice of this convolution product
may be arbitrary. Nevertheless, in the mathematical physics context, often there are cases where such choice is
forced by some {'ad hoc'} physical reasons (as, for example, causality $-$ see [Vr]).} product $\ast$ (of \it
convolution\rm) of elements of $\langle X\rangle$. The resulting algebraic structure\footnote{The unitary ring
(in general, non-commutative) of this structure $\mathcal{C}_A(X)$, is $(\langle X\rangle,+,\ast)$, whereas the
convolution product $\ast$ and the $A$-module product $\cdot$, are linked by the compatibility relation
$a\cdot(x\ast y)=(a\cdot x)\ast y=x\ast(a\cdot y)\ \ \forall a\in A,\ \forall x,y\in\langle X\rangle$; as usual,
the product $\cdot$ is implicit.} ${\cal C}_A(X)=(\langle X\rangle,+,\cdot,\ast)$, is said to be a \it
convolution structure \rm associated to the free $A$-module ${\cal M}_A(X)$.\\It is important to point out that
such structure is strictly related to the choice of the product $\ast$, and, to this purpose, the following
remarks are meaningful. When the set $X$ is already endowed with a given algebraic structure of the type
$(X,\diamond)$ (for example, that of EBB-groupoid), formal coherence principles imposes that such (convolution)
product must be 'predetermined' by such preexistent structure. For instance, it is usually required such product
to be the result of the linear $A$-extension of the operation $\diamond$, eventually taking into account further
informal requirements (of physical nature).\\In particular, in view of the next arguments, $A$ will be a scalar
field $\mathbb{K}$, so that the $A$-module $(\langle X\rangle,+,\cdot)$ is a $\mathbb{K}$-linear space that will
be extended to a linear $\mathbb{K}$-algebra (often said the \it convolution $\mathbb{K}$-algebra \rm of the
given $A$-module), by means a convolution product.\\Historically, the first structures of convolution
$\mathbb{K}$-algebras were the so-called \it {group algebra}, \rm convolution structures made on a given group.
Group algebras were introduced (for finite groups) by T.Molien and G. Frobenius, for investigations of
representations of these groups. Subsequently, at the beginning of the 20{th}-century, I. Schur ([Sh]) and H.
Weyl ([We]) used systematically the group algebras for investigations of compact groups and commutative locally
compact groups.\\For applications of group algebra structures to quantum mechanics, see, for example, [Lo].
\subsection*{3. The notion of EBB-groupoid algebra} With any EBB-groupoid, a unique, well defined,
natural structure of linear $\mathbb{K}$-algebra is associated by means a precise convolution product $\ast$.\\
If $\mathbb{K}$ is a field, let us consider the free $\mathbb{K}$-module generated by the support of the given
EBB-groupoid, hence we will consider the related convolution structure provided by an adapted convolution
product. Such a group algebra is more appropriately called the \it EBB-groupoid algebra \rm associated to the
given EBB-groupoid (although it is nothing else that a group algebra on the support of a EBB-groupoid).\\ If
${\cal G}=(G,G^{(0)},r,s,\star)$ is a EBB-groupoid and $\mathbb{K}$ is a scalar field, we put $X=G$, and thus
$\langle G\rangle=\{\sum_{g\in G}\lambda(g)g;\ \lambda(g)\in\mathbb{K},\ \forall g\in G\}$ is the set of the
formal combinations (with coefficients in $\mathbb{K}$) of elements of $G$, equipped with the operations
$$\big(\sum_{g\in G}\lambda(g)g\big)+\big(\sum_{g\in G}\mu(g)g\big)=\sum_{g\in G}(\lambda(g)+\mu(g))g,\leqno(1)$$
$$\mu\cdot\big(\sum_{g\in G}\lambda(g)g\big)=\sum_{g\in G}(\mu\lambda(g))g\ \ \ \forall\mu\in\mathbb{K}.\leqno(2)$$
${\cal M}_{\mathbb{K}}(G)=(\langle G\rangle,+,\cdot)$ is a free $\mathbb{K}$-module (that is a
$\mathbb{K}$-linear space) with base $G$. We define the convolution product$$\big(\sum_{g\in
G}\lambda(g)g\big)\ast\big(\sum_{g\in G}\mu(g)g\big)=\sum_{g\in G}\xi(g)g\leqno(3)$$where (with
Cauchy)$$\xi(g)=\sum_{g_1\star g_2=g}\lambda(g_1)\mu(g_2)\ \ \ \forall g\in G.\leqno(4)$$Therefore, it is
immediate to verify that ${\cal C}_{\mathbb{K}}(G)=(\langle G\rangle,+,\cdot,\ast)$, as convolution structure
associated to ${\cal M}(G)$, is a linear $\mathbb{K}$-algebra, that we will call the \it EBB-groupoid algebra
\rm (over the field $\mathbb{K}$) associated to the EBB-groupoid ${\cal G}=(G,G^{(0)},r,s,\star)$. We will
denote it by ${\cal A}_{\mathbb{K}}({\cal G})$. As regards what has been said at the end of § 2, it is possible
a (unique) linear extension of the algebraic relations of the EBB-groupoid $\cal G$ to ${\cal
A}_{\mathbb{K}}(\cal G)$.\subsection*{4. Brief outlines of atomic spectroscopy\footnote{See [Em], Chapt. 7, and
[An], [Ec], [Fe], [Gi], [Hu], [Lu].}}The physical phenomenology leading to the first theoretical formulations of
quantum mechanics, were the atomic spectroscopy of emission and absorption of electromagnetic waves.
Nevertheless, the classical physics failed in the interpretation of the structure of the observed atomic spectra
of the various chemical elements. An atomic spectrum of emission [absorption ] of an arbitrary chemical element,
is characterized by a well defined sequence of spectral lines. Each spectroscopic line is related to the
emission [absorption] of an electromagnetic radiation of well precise frequency, from the atom of the given
chemical element under examination. Hence, each spectral line is identified by that well-determined frequency
$\nu$ of the e.m. radiation corresponding to it. Moreover, such lines appears organized into groups called  (\it
spectral\rm) \it series \rm (of the given spectrum).\\From these experimental investigations, born the atomic
spectroscopy of the beginning of the 20{th}-century. It were, mainly, a coherent set of qualitative and
descriptive rules concerning the symmetry and regularity properties of these spectral lines. The new quantum
theory was built on the basis of it.\\The first semi-quantitative spectroscopic rules has been formulated with
the quantum theory of N.H. Bohr, A. Sommerfeld and W. Wilson in the years 1913-1914, on the basis of the
pioneering works of M. Planck, A. Einstein, E. Rutherford, and others.\\\\Nevertheless, in experimental
spectroscopy, the works of J.R. Rydberg ({see} {[38]}) and W. Ritz ({see
[38]}) got a prominent rule, and led to the formulation of their homonymous principle.\\
From the analysis of the (spectral) series of lines of several single atomic spectra, Rydberg established the
following (\it {Rydberg}\rm) \it {combination principle}\rm:\begin{description}\item $\bullet_1$ \ each spectral
lines, of any series, can be described in terms of suitable \it spectral terms\ \ \rm $T_{(s)}(n)\
s,n\in\mathbb{N}$;\item $\bullet_2$ \ \ the frequency (or the wave number) of a line is given by a relation of
the type$$\nu_{(n,n')}^{(s,s')}=T_{(s')}(n')-T_{(s)}(n)\leqno(\spadesuit)$$ where $s,s'$ are indices of spectral
series (of lines), $n,n'$ indices of lines (of spectral series), and $\lim_{n\rightarrow\infty}T_{(s)}(n)=0$ for
any fixed $s$;\item $\bullet_3$ \ \ there are precise \it selection rules \rm that are imposed
instructions\footnote{For instance, a simple selection rule is $|s-s'|=1$. In any case, according to these
rules, not every possible combination of values of $s,s'$, corresponds to an effectively observed spectral
lines, but only those that satisfy well defined algebraic relations.} on the possible values $s,s'$ of
$(\spadesuit)$ in order that the frequency predicted by this formula represents observed spectral
lines.\end{description} Successively, Ritz stated that:
\begin{description}\item 1) \ \ on the basis of the \it {Rydberg's formula}, \rm found
for the hydrogenoid atoms, according to$$T_{(s)}(n)={R_H}/{(n-a_s)^2}$$(where $R_H$ is a universal constant (of
Rydberg) and $a_0,a_1,a_2,...$ are real constants, typical of each hydrogenoid atom, and that characterizes the
different spectral series of its spectrum), it is possible to formulate, for the spectral terms, the following
more general expression$$T_{(s)}(n)=\Phi_{(s)}(n)(K/n^2)\qquad\mbox{\rm(\it {Ritz-Rydberg formula}\rm)}$$where
$K$ is a universal constant, and $\lim_{n\rightarrow\infty}\Phi_{(s)}(n)=1$ for each fixed $s$;\item 2) \ \ the
frequencies of the lines of each spectral series (that is, for each fixed $s'=s$), verify the following \it
Ritz's combination \rm(or \it composition\rm) \it law \rm
$\nu_{(i,j)}^{(s)}+\nu_{(j,k)}^{(s)}=\nu_{(i,k)}^{(s)}$.\end{description}Subsequently, Ritz verified the
validity of law 2) also for the lines of different series, reaching to a more general \it Ritz-Rydberg
composition law\footnote{Independently by eventual selection rules on $s,s'$, as, instead, required by the
Rydberg's combination principle (see $\bullet_3$). Moreover, $(\clubsuit)$ is verified by $(\spadesuit)$, under
the suitable selection rules.}:$$\nu_{(i,j)}^{(s,s')}+\nu_{(j,k)}^{(s,s')}=\nu_{(i,k)}^{(s,s')}\ \ \ \ \ \forall
s,s',\ \ \ \ \forall i,j,k.\leqno(\clubsuit)$$\rm The principle achieved by the above conditions 1) and 2), with
the last extension due to Ritz, it is usually called the \it Ritz-Rydberg composition principle \rm(as extension
of the Rydberg's composition principle). \rm The main interesting part of this principle is the Ritz-Rydberg
composition law \footnote{Neglecting the series indices $(s,s')$.} $\nu_{(i,j)}+\nu_{(j,k)}=\nu_{(i,k)}$.\\ The
principle has been confirmed by a large class of spectroscopic phenomena, ranging from atomic spectra to the
molecular ones, from those optical to the X ray spectra, and so on.\\\\ \rm The class of the fundamental
Franck-Hertz experiments ([FH]; [CCP], Chap. IV, § 10; [Her]), experimentally confirmed the existence of
discrete energy levels in an atom (predicted by N.H. Bohr), each characterized by a certain frequency $\nu$
($=E/h$, $E$ being the typical energy value of the given level). Hence, the set of the energy levels of an atom
is characterized by a well determined set of frequencies ${\cal F}_I=\{\nu_i; \nu_i\in\mathbb{R}^+,\ i\in
I\subseteq\mathbb{N}\}\subseteq\mathbb{R}^+$, that the Franck-Hertz experiences proved to be finite or
countable, hence labelled by a subset $I$ of $\mathbb{N}$. Since $I$ is a non-empty subset of
$(\mathbb{N};\leq)$ (with the usual order), ${\cal F}_I$ can be ordered according to the increasing indices of
$I$, so that we can assume that the map $\rho:i\rightarrow\nu_i\in{\cal F}_I\ \ \forall i\in I$ is monotonically
increasing and thus $I\cong{\cal F}_I$. Moreover, at each energy level correspond the energy $E_i=h\nu_i$, where
$h$ is the Planck's constant. Furthermore, we recall another basic principle of the spectroscopy, namely the \it
principle of A. Conway \rm (see [Mo], Chapt. 13, § 11), according to any spectral line is produced, once at a
time, by a unique (perinuclear) electron, with $\nu_{ij}\neq 0$ whenever $i\neq j$.\\\\A spectral line
correspond to the electromagnetic radiation involved in the transition of an electron from a given level to
another. If, throughout such transition $i\rightarrow j$, the initial energy level is $\nu_i$ and the final is
$\nu_j$, with $i\neq j$, then the only \it physical observable \rm(according to P.A.M. Dirac $-$ see [Di2]) is
the electromagnetic radiation of emission (if $i>j$), or of absorption (if $i<j$), with frequency
$\nu_{i\rightarrow j}=\nu_j-\nu_i=(E_j-E_i)/h$. In the emission it is $\nu_{i\rightarrow j}<0$, whereas in the
absorption it is $\nu_{i\rightarrow j}>0$. Let we put\footnote{With $\nu_{(i,j)}=\nu_{ij}$.} $\nu_{i\rightarrow
j}=\nu_{ij}$, so that, if $\nu_{ij}=\nu_j-\nu_i<0$, then we have an emission line, whereas, if
$\nu_{ji}=-\nu_{ij}>0$, then we have an absorption line. It follows the existence of a simple (opposite)
symmetry, called the \it Kirchhoff-Bunsen \rm(\it inversion\rm) \it symmetry \rm(or \it KB-symmetry\rm), between
the emission spectrum and the absorption spectrum of the same atom, given by $\nu_{ij}=-\nu_{ji}$ (and that
follows from the Ritz-Rydberg composition law). Therefore, the lines of the atomic spectrum (of
emission/absorption) of a chemical element, are represented (in $\mathbb{R}$) by the elements of the set
$\Delta{\cal F}_I=\{\nu_{ij}=\nu_j-\nu_i;\ i,j\in I\subseteq\mathbb{N}\}\subseteq\mathbb{R}$, symmetric respect
to the origin, whose positive part represents the absorption spectrum, whereas that negative represents the
emission spectrum.\\ Finally, other spectroscopic principles prescribes symmetries and regularities of this set,
from which it is possible infer further algebraic properties of it. However, as seen, between these principles,
the Ritz-Rydberg combination principle has a prominent rule in justifies the intrinsic non-commutativity of the
formal quantum theory.\subsection*{5. The Heisenberg-Born-Jordan EBB-groupoid}In general, $(\Delta{\cal F}_I;+)$
is not a subgroup of the commutative group $(\mathbb{R};+)$ because, if $\nu_{ij},\nu_{lk}\in\Delta{\cal F}_I$,
not even $\nu_{ij}\pm\nu_{lk}$ correspond to an observed spectral line, that is may be
$\nu_{ij}\pm\nu_{lk}\notin\Delta{\cal F}_I$. The elements of $\Delta{\cal F}_I$ combines by means the
Ritz-Rydberg combination principle, according to $\nu_{ij}+\nu_{lk}\in\Delta{\cal F}_I$ if and only
if\footnote{This last condition suggests the formal presence of a EBB-groupoid structure. Indeed, this structure
is endowed with a partial binary operation defined on a domain $G^{(2)}=\{(g_1,g_2)\in G\times G;
s(g_1)=r(g_2)\}$, where the condition $s(g_1)=r(g_2)$ correspond to $j=l$, as we shall see.} $j=l$, whence
$\nu_{ij}+\nu_{lk}=\nu_{ik}$. It is just this last principle that make $+$ a partial law\footnote{That is, do
not defined for all possible pairs $(\nu_{ij},\nu_{lk})$.} in $\Delta{\cal F}_I$, so that $(\Delta{\cal F}_I;+)$
not only is not a subgroup, but neither a groupoid in the sense of Universal Algebra (see [Co], or\footnote{This
Author introduces a structure that he call \it {magma}, \rm corresponding to the notion of groupoid of Universal
Algebra.} [Bou]).\\ We will prove, instead, that it is a EBB-groupoid. Indeed, if $G=\Delta{\cal F}_I,$ we have
$G^{(0)}={\cal F}_I,$ $r(\nu_{ij})=\nu_i,$ and $s(\nu_{ij})=\nu_j$, with $\nu_{ij}\star\nu_{lk}$ defined on
$G^{(2)}=\{(\nu_{ij},\nu_{lk})\in\Delta{\cal F}_I\times\Delta{\cal F}_I;\ s(\nu_{ij})=r(\nu_{lk})\}$. Therefore,
since $s(\nu_{ij})=\nu_j=\nu_l=r(\nu_{lk})$, we have $\nu_j=\nu_l$, from which (by the bijectivity of
$\rho:i\rightarrow\nu_i$ of § 4) follows that $j=l$, hence $\nu_{ij}\star\nu_{lk}=\nu_{ij}\star\nu_{jk}$. Thus,
if $\star=\tilde{+}$, we have\footnote{$\tilde{+}$ denotes the usual addition of $(\mathbb{R};+)$, only
partially defined on $G^{(2)}\subset\mathbb{R}^2$, and that, substantially, represents the Ritz-Rydberg
composition law.} $\nu_{ij}\star\nu_{jk}=\nu_{ij}\tilde{+}\nu_{jk}=\nu_{ik}$ by the Ritz-Rydberg composition
law, and hence it is immediate to verify {that} ${\cal G}_{HBJ}({\cal F}_I)=(\Delta{\cal F}_I,{\cal
F}_I,r,s,\tilde{+})$ is a EBB-groupoid, said to be\footnote{The motivation for this terminology will be given at
§ 6.} the \it Heisenberg-Born-Jordan EBB-groupoid \rm(or \it HBJ EBB-groupoid\rm).\\There are many different
representation (or equivalent models) of ${\cal G}_{HBJ}({\cal F}_I)$. We will consider a first representation
in $\mathbb{N}^2$. As regard what has been said above, the map $\xi:\nu_{ij}\rightarrow (i,j)\in
I^2(\subseteq\mathbb{N}^2)\ \ \forall\nu_{ij}\in \Delta{\cal F}_I$ is bijective, so that $\Delta{\cal F}_I\cong
I^2$, that is $I^2$ is a Cartesian representation of $\Delta{\cal F}_I$. Considering the Cartesian lattice
$\mathbb{N}^2$, if $\Delta(\mathbb{N}^2)=\{(i,i);\ i\in\mathbb{N}\}$, then the points $(i,j)\in I^2$ with $i<j\
[i>j]$ represents the lines of the absorption [emission] spectrum, whereas the points with $i=j$ represents the
energy levels of the atom because, being $\rho^{-1}:\nu_i\rightarrow i\in I\ \ \forall\nu_i\in{\cal F}_I$
bijective and $\alpha:i\rightarrow(i,i)$ bijection of $\mathbb{N}$ in $\Delta(\mathbb{N}^2)$, we have that
$(\alpha\circ\rho^{-1})({\cal F}_I)=\{(i,i);\ i\in I)\}=\Delta(I^2)(\subseteq\Delta(\mathbb{N}^2))$ represents
${\cal F}_I$ in $\mathbb{N}^2$.\\ Thus, we have a Cartesian representation (in $\mathbb{N}^2$) of the atomic
spectrum $\Delta{\cal F}_I$: the upper half plane respect to $\Delta(\mathbb{N}^2)$ represents the absorption
spectrum, whereas the lower half plane represents the emission spectrum, and finally $\Delta(I^2)$ represents
the set of energy levels of the atom. The points of the two emission and absorption half planes, are pairwise
correlated by the (inversion) Kirchhoff-Bunsen symmetry, that it is a simple reflexive symmetry\footnote{From
this symmetry it is possible to construct a second equivalent representation of the given EBB-groupoid.} respect
to $\Delta(\mathbb{N}^2)$.\\ The map $\xi^{-1}:(i,j)(\in I^2)\rightarrow\nu_{ij}$ provides the kinematical time
evolution of a dynamic system through the coordinate $\mathfrak{q}$ and the momentum $\mathfrak{p}$ given by the
(Hermitian) matrices$$\mathfrak{q}=[q_{ij}\ e^{2\pi i\nu_{ij}t}],\ \ \mathfrak{p}=[p_{ij}\ e^{2\pi
i\nu_{ij}t}]\qquad\mbox{\rm(\it Heisenberg's\ representation\rm)},\leqno(\maltese)$$from which follows that any
other physical observable $\mathfrak{g}(\mathfrak{q},\mathfrak{p})$ can always be written in the form
$\mathfrak{g}=[{g_{ij}}\ e^{2\pi i\nu_{ij}t}]$. The celebrated canonical commutation relations between
$\mathfrak{q}$ and $\mathfrak{p}$, does follows from the Kuhn-Thomas relation (interpreted according to
Perturbation Theory; see [BJ]).
\\\\Finally, we have the identifications $\Delta{\cal F}_I\cong I^2\cong{\cal F}_I^2$, that give a
representation (in $\mathbb{N}^2$) of the algebraic system ${\cal G}_{HBJ}({\cal F}_I)$. In fact, if we put
$X=I$, it is immediate to verify that ${\cal G}_{HBJ}({\cal F}_I)$ is identifiable (since isomorphic to it) with
the natural EBB-groupoid ${\cal G}_{Br}(X)$ (of the last part of § 1) with $X=I$, $G=I^2,\ G^{(0)}=\Delta(I^2),\
r:(i,j)\rightarrow i,\ s:(i,j)\rightarrow j,\ (i,j)\star(j,k)=(i,k)$;
${\cal G}_{Br}(I)$ is said to be the \it Heisenberg-Born-Jordan natural EBB-groupoid.\\
\rm Two EBB-groupoids ${\cal G}_1=(G_1,G^{(0)}_1,r_1,s_1,\star_1),\ {\cal G}_2=(G_2,G^{(0)}_2,r_2,s_2,\star_2)$
are said\footnote{This definition of isomorphism must be intended in the sense of category theory.} \it
isomorphic \rm if there exists $\psi:G_1\rightarrow G_2,\ \psi_0:G^{(0)}_1\rightarrow G^{(0)}_2$ both bijective,
such that $r_2\circ\psi=\psi_0\circ r_1,\ s_2\circ\psi=s_1\circ\psi_0$ and $\psi(g\star_1
g')=\psi(g)\star_2\psi(g')\ \ \forall g,g'\in G_1$. In this case, we will write ${\cal G}_1\cong{\cal G}_2$.\\
If ${\cal G}_1={\cal G}_{HBJ}({\cal F}_I),\ {\cal G}_2={\cal G}_{Br}(I)$, the maps
$\psi:\nu_{ij}\rightarrow(i,j)\in I^2 \ \forall\nu_{ij}\in\Delta{\cal F}_I$,
$\psi_0:\nu_i\rightarrow(i,i)\in\Delta(I^2)\ \ \forall\nu_i\in{\cal F}_I$, are bijective and make isomorphic the
two given B-groupoid\footnote{For instance, we have
$\psi(\nu_{ij}\tilde{+}\nu_{jk})=\psi(\nu_{ik})=(i,k)=(i,j)\star(j,k)=\psi(\nu_{ij})\star\psi(\nu_{jk})$, the
remaining conditions being easily verified.}, that is ${\cal G}_{HBJ}({\cal F}_I)\cong{\cal G}_{Br}(I)$, hence
they are identifiable.\subsection*{6. The Heisenberg-Born-Jordan EBB-groupoid algebra}Let us prove that the
EBB-groupoid algebra of the Heisenberg-Born-Jordan EBB-groupoid, in general, is a non-commutative linear
$\mathbb{K}$-algebra, isomorphic, in the finite dimensional case, to a well defined matrix $\mathbb{K}$-algebra.
\\First, from ${\cal G}_{Br}(I)\cong{\cal G}_{HBJ}({\cal F}_I)$, follows that\footnote{Indeed, the base of the
free $\mathbb{K}$-modules constructed on such EBB-groupoid, that is to say their supports, are respectively
$I^2$ and $\Delta{\cal F}_I$, and since $I^2\cong\Delta{\cal F}_I$, follows that the respective
$\mathbb{K}$-modules are $\mathbb{K}$-isomorphic (see § 3), whence ${\cal A}_{\mathbb{K}}({\cal
G}_{Br}(I))\cong{\cal A}_{\mathbb{K}}({\cal G}_{HBJ}({\cal F}_I))$.} ${\cal A}_{\mathbb{K}}({\cal
G}_{Br}(I))\cong{\cal A}_{\mathbb{K}}({\cal G}_{HBJ}({\cal F}_I))$. These two isomorphic structures represent
the so called\footnote{See § 7.} \it {Heisenberg-Born}-Jordan EBB-groupoid algebra \rm(or \it HBJ
EBB-algebra\rm).\\\\It is ${\cal A}_{\mathbb{K}}({\cal G}_{HBJ}({\cal F}_I))=(\langle\Delta{\cal
F}_I\rangle,+,\cdot,\ast)$ with$$\langle\Delta{\cal F}_I\rangle=\{\sum_{g\in\Delta{\cal F}_I}\lambda(g)g;\
\lambda(g)\in\mathbb{K},\ g\in\Delta{\cal F}_I\}=$$
$$=\{\sum_{\nu_{ij}\in\Delta{\cal F}_I}\lambda(\nu_{ij})\nu_{ij};\ \lambda(\nu_{ij})\in\mathbb{K},\
\nu_{ij}\in\Delta{\cal F}_I\},$$the operations $+,\cdot$ being definite as in (1), (2), whereas the convolution
product $\ast$ is defined as follow. If we consider two arbitrary elements of $\langle\Delta{\cal F}_I\rangle$,
their convolution product must be an element of this set, and hence it is expressible as unique linear
combination of the base elements of $\Delta{\cal F}_I$. Hence, in general, we
have$$\big(\sum_{\nu_{ij}\in\Delta{\cal F}_I}\lambda(\nu_{ij})\nu_{ij}\big)\ast\big(\sum_{\nu_{lk}\in\Delta{\cal
F}_I}\mu(\nu_{lk})\nu_{lk}\big)=\sum_{\nu_{ps}\in\Delta{\cal F}_I}\xi(\nu_{ps})\nu_{ps},$$where
$\xi(\nu_{ps})\in\mathbb{K}$ are uniquely determined by:
$$\xi(\nu_{ps})\doteq\sum_{\nu_{ij}\tilde{+}\nu_{lk}=\nu_{ps}}\lambda(\nu_{ij})\mu(\nu_{lk}),$$ and, since
$\nu_{ij}\tilde{+}\nu_{lk}$ is defined if and only if $s(\nu_{ij})=r(\nu_{lk})$, that is $j=l$, we have
$\nu_{ij}\tilde{+}\nu_{lk}=\nu_{ij}\tilde{+}\nu_{jk}=\nu_{ik}$, and thus $\nu_{ps}=\nu_{ik}$, that is $p=i,s=k$,
so that
$$\xi(\nu_{ik})=\sum_{\nu_{ij}\tilde{+}\nu_{jk}=\nu_{ik}}\lambda(\nu_{ij})\mu(\nu_{jk}).\leqno(\diamondsuit)$$In
the last sum, the indexes $i,k$ are saturated while $j(\in I)$ is free, so that the sum is extended only to last
index, that is we can write\footnote{All these convolution structures are based on the Rytz-Rydberg composition
law, and, the latter, has been at the foundations of the works of W. Heisenberg, M. Born and P. Jordan on matrix
mechanics (see [An], [BJ], [Di1], [Di2], [Ec], [Fe], [Gi], [HBJ], [He1,2], [Hu], [Lu], [Sch]). For modern
quantum theories see [St], [Em], whereas for the geometric developments of the non-commutativity, and its
physical applications, see [Co1], [Co2].}$$\xi(\nu_{ik})=\sum_{j\in
I}\lambda(\nu_{ij})\mu(\nu_{jk})\leqno(\heartsuit)$$and this is the expression of the generic element of the
product\footnote{Rows by columns.} matrix of the two formal matrices\footnote{Of order $card\ I$.}
$[\lambda(\nu_{ij})],[\mu(\nu_{lk})]$. Then, in general, the Heisenberg-Born-Jordan EBB-groupoid algebra ${\cal
A}_{\mathbb{K}}({\cal G}_{HBJ}({\cal F}_{I}))$, is non-commutative, and, if card $I<\infty$, then it is
isomorphic to a matrix $\mathbb{K}$-algebra (see [Pi], Chap. 5, Theor. 3-2, or [Bou], [Ma],
[Mac]).\\Nevertheless, since, in general, $I$ is not finite, it follows that
$\mathcal{A}_{\mathbb{K}}(\mathcal{G}_{HBJ}(\mathcal{F}_I))$ is not a finitely generated
$\mathbb{K}$-algebra.\subsection*{7. The notion of Hopf algebra\footnote{We follow [Ka], [KS], [CP], [Md], and
[Mn], [As].}}Let $V_{\mathbb{K}}$ be a $\mathbb{K}$-linear space and id its identity map. A \it
$\mathbb{K}$-algebra \rm is an algebraic system ${\cal A}=(V_{\mathbb{K}},m,\eta)$ with $m:V\otimes V\rightarrow
V$ (\it product\rm) and $\eta:\mathbb{K}\rightarrow V$ (\it unit\rm) $\mathbb{K}$-linear maps such that: 1) (\it
associativity\rm) $m\circ(m\otimes\rm id)=\it m\circ(\rm id\otimes\it m)$; 2) $m\circ(\eta\otimes\rm id)=\it
m\circ(\rm id\otimes\eta)=id$. We have $m(a\otimes b)=ab$. \\A \it $\mathbb{K}$-coalgebra \rm is an algebraic
system $c{\cal A}=(V_{\mathbb{K}},\Delta,\varepsilon)$ with $\Delta:V\rightarrow V\otimes V$ (\it coproduct\rm)
and $\varepsilon:V\rightarrow\mathbb{K}$ (\it counit\rm) $\mathbb{K}$-linear maps such that: 1') (\it
coassociativity\rm) $(\Delta\otimes \rm id)\circ\Delta=(id\otimes\Delta)\circ\Delta$; 2')
$(\varepsilon\otimes\rm id)\circ\Delta=(id\otimes\varepsilon)\circ\Delta=id$. A \it $\mathbb{K}$-bialgebra \rm
is an algebraic system $b{\cal A}=(V_{\mathbb{K}},\Delta,\varepsilon,m,\eta)$ such that
$(V_{\mathbb{K}},\Delta,\varepsilon)$ is a $\mathbb{K}$-coalgebra, $(V_{\mathbb{K}},m,\eta)$ a
$\mathbb{K}$-algebra and $\Delta,\varepsilon$ $\mathbb{K}$-algebras homomorphism (see [Ka], Chapt. III, Theor.
III.2.1). Let $f\in End(V_{\mathbb{K}})$, and $\{v_i;i\in I\}$ a base of $V_{\mathbb{K}}$ with
dim$_{\mathbb{K}}V_{\mathbb{K}}=card\ I$; then $\{v_i\otimes v_j;(i,j)\in I^2\}$ is a base of $V\otimes V$, so
that $v=\sum_{(i,j)\in I_v}\lambda_{ij}(v)v_i\otimes v_j$ for certain $\lambda_{ij}(v)\in\mathbb{K}$ and
$I_v\subseteq I^2$, for any $v\in V\otimes V$. Then, we can use the following  \rm(\it Sweedler\rm) \it sigma
notation \rm according to we will write simply $v=\sum_{(v)}v'\otimes v''$ for given $v',v''\in V_{\mathbb{K}}$;
in particular, since $\Delta(v)\in V\otimes V\ \ \forall v\in V_{\mathbb{K}}$, we will have
$\Delta(v)=\sum_{(v)}v'\otimes v''$. If $(V_{\mathbb{K}},\Delta,\varepsilon)$ is a $\mathbb{K}$-coalgebra and
$(W_{\mathbb{K}},m,\eta)$ a $\mathbb{K}$-algebra, for each $f,g\in Hom(V,W)$ we have $f\otimes g\in Hom(V\otimes
V,W\otimes W)$, so that, by exploiting the relations$$V\stackrel{\Delta}{\longrightarrow}V\otimes
V\stackrel{f\otimes g}{\longrightarrow}W\otimes W\stackrel{m}{\longrightarrow}W,$$it is possible to consider the
composition map $m\circ(f\otimes g)\circ\Delta\in Hom(V,W)$, denoted by $f\ \hat{\ast}\ g$ and given by$$(f\
\hat{\ast}\ g)(v)=(m\circ(f\otimes g)\circ\Delta)(v)=(m\circ(f\otimes g))(\Delta(x))=$$
$$=(m\circ(f\otimes g))(\sum_{(v)}v'\otimes v'')=m(\sum_{(v)}(f\otimes g)(v'\otimes v''))=$$
$$=m(\sum_{(v)}f(v')\otimes g(v''))=\sum_{(v)}f(v')g(v'')\ \ \forall v\in V,$$defining the (internal) binary
operation (also called \it convolution\rm)
$$\hat{\ast}:Hom(V,W)\times Hom(V,W)\rightarrow Hom(V,W)$$ $$(f,g)\rightsquigarrow f\ \hat{\ast}\ g\ \ \
\forall f,g\in Hom(V,W).$$If $b{\cal A}=(V_{\mathbb{K}},\Delta,\varepsilon,m,\eta)$ is a $\mathbb{K}$-bialgebra,
what said above also subsists if we put $V=W$; an element $a\in End(V)$ is said to be an \it antipode \rm of
$b\cal A$ if $a\ \hat{\ast}\ \rm id=id\ \hat{\ast}\ \it a=\eta\circ\varepsilon$, and the algebraic system
$(b{\cal A},a)$ is called a \it Hopf \ $\mathbb{K}$-algebra \!\footnote{In a $\mathbb{K}$-bialgebra, may exist,
at most, only one antipode (see [Ka], Def. III.3.2., p. 51).} \rm of support $V_{K}$.\\ The commutativity (or
not) [cocommutativity (or not)] of such a structure, follows from the commutativity (or not) [cocommutativity
(or not)] of the linear $\mathbb{K}$-algebra $(V_{\mathbb{K}},m,\eta)$ [$\mathbb{K}$-coalgebra
$(V_{\mathbb{K}},\Delta,\varepsilon)$].\\A $\mathbb{K}$-bialgebra
$b\mathcal{A}=(V_{\mathbb{K}},\Delta,\varepsilon,m,\eta)$ is said to be \it quasi-cocommutative \rm if there
exists an invertible element $R$ (called an \it universal R-matrix\rm) of $V_{\mathbb{K}}\otimes V_{\mathbb{K}}$
such that $\Delta^{op}(x)=R\Delta(x)R^{-1}\ \ \forall x\in V_{\mathbb{K}}$, where
$\Delta^{op}=\tau_{V_{\mathbb{K}},V_{\mathbb{K}}}\circ\Delta$ is the opposite coproduct on $V_{\mathbb{K}}$ and
$\tau_{V_{\mathbb{K}},V_{\mathbb{K}}}$ is the flip switching the factors. We will denote such a
quasi-cocommutative $\mathbb{K}$-bialgebra with $(V_{\mathbb{K}},\Delta,\varepsilon,m,\eta,R)$.\\Any
cocommutative $\mathbb{K}$-bialgebra is also quasi-cocommutative with universal R-matrix
$R=1_{V_{\mathbb{K}}}\otimes 1_{V_{\mathbb{K}}}$. A Hopf $\mathbb{K}$-algebra whose underling
$\mathbb{K}$-bialgebra has a universal R-matrix, is said to be a \it quasi-cocommutative Hopf\ \
$\mathbb{K}$-algebra.\\ \rm For any $R\in V_{\mathbb{K}}\otimes V_{\mathbb{K}}$, we set $R_{12}=R\otimes
1_{V_{\mathbb{K}}}\in V_{\mathbb{K}}\otimes V_{\mathbb{K}}\otimes V_{\mathbb{K}}=\bigotimes^3V_{\mathbb{K}}$,
$R_{23}=1_{V_{\mathbb{K}}}\otimes R\in\bigotimes^3V_{\mathbb{K}}$, $R_{13}=(\mbox{\rm
id}\otimes\tau_{V_{\mathbb{K}},V_{\mathbb{K}}})(R_{12})=(\tau_{V_{\mathbb{K}},V_{\mathbb{K}}}\otimes\mbox{\rm
id})(R_{23})\in\bigotimes^3V_{\mathbb{K}}$. A quasi-cocommutative $\mathbb{K}$-bialgebra [Hopf
$\mathbb{K}$-algebra] $b\mathcal{A}=(V_{\mathbb{K}},\Delta,\varepsilon,m,\eta,R)$ [($b\mathcal{A},a)$] is \it
braided \rm(or \it quasi-triangular\rm) if the universal R-matrix $R$ satisfies the relations
$(\Delta\otimes\mbox{\rm id}_{V_{\mathbb{K}}})(R)=R_{13}R_{23}$ and $(\mbox{\rm
id}_{V_{\mathbb{K}}}\otimes\Delta)(R)=R_{13}R_{12}$. All cocommutative bialgebras are braided with universal
R-matrix $R=1_{V_{\mathbb{K}}}\otimes 1_{V_{\mathbb{K}}}$.\\Often, a non-cocommutative braided Hopf
$\mathbb{K}$-algebra is called a \it quantum group.\rm\subsection*{8. The HBJ EBBH-algebra}It is possible to
associate a natural structure of (non-commutative) braided Hopf $\mathbb{K}$-algebra to the HBJ EBB-algebra. So,
we will get an explicit example of non-commutative Hopf $\mathbb{K}$-algebra, that can be taken as basic
structure of a (particular) quantum group.\\\\On ${\cal A}_{\mathbb{K}}({\cal G}_{HBJ}({\cal
F}_I))=(\langle\Delta{\cal F}_I\rangle,+,\cdot,\ast)$, let us consider the unique (see § 2) linear extension
$\tilde{i}$ of the inversion map $i_G:g\rightarrow g^{-1}$ (with $G=\Delta{\cal F}_I$) to $\langle\Delta{\cal
F}_I\rangle$ and set\footnote{In the notations of § 7.} (if necessary, in the tensor product algebra ${\cal
A}_{\mathbb{K}}\otimes{\cal A}_{\mathbb{K}}$): $V_{\mathbb{K}}=(\langle\Delta{\cal F}_I\rangle,+,\cdot)$,
$m=\ast,\ \eta=\eta(1)(=1\ \mbox{the\ unit\ of}\ {\cal A}_{\mathbb{K}}({\cal G}_{HBJ}({\cal F}_I))),\
\Delta(x)=x\otimes x\ \mbox{(group-like\ elements)},\ \varepsilon(x)=1\ \ \forall x\in\langle\Delta{\cal
F}_I\rangle$ and $a=\tilde{i}\in End(\langle\Delta{\cal F}_I\rangle)$. Hence$$\Delta(m(x,y))=\Delta(x\ast
y)=(x\ast y)\otimes(x\ast y)=\footnote{See [Ka], Chapt. II, § II.4.}$$
$$=(x\otimes y)\ast(x\otimes y)=\Delta(x)\ast\Delta(y)=m(\Delta(x),\Delta(y))\ \ \forall
x,y\in\langle\Delta({\cal F}_I)\rangle,$$
$$\varepsilon(m(x,y))=\varepsilon(x\ast y)=\varepsilon(z)=1=1\ast
1=$$ $$=\varepsilon(x)\ast\varepsilon(y)=m(\varepsilon(x),\varepsilon(y))\ \ \forall x,y\in\langle\Delta({\cal
F}_I)\rangle$$(where $z=m(x,y)$, and $\varepsilon(x)=1\ \forall x\in\Delta({\cal F}_I)$), so that
$\Delta,\varepsilon$ are homomorphism of $\mathbb{K}$-algebras.\\ Therefore, it is immediate to prove that
$b{\cal A}_{\mathbb{K}}({\cal G}_{HBJ}({\cal F}_I))=(\langle\Delta{\cal F}_I\rangle,\Delta,\varepsilon,m,\eta)$
is a (non-commutative) $\mathbb{K}$-bialgebra (see, also, [KS], Chapt. I, § 1.2.6., Example 7). In fact, we
have$$(a\ \hat{\ast}\
\mbox{id})(x)=(m\circ(\tilde{i}\otimes\mbox{id})\circ\Delta)(x)=m(\tilde{i}\otimes\mbox{id})(x\otimes x)=
=m(\tilde{i}(x)\otimes x)=$$ $$=m(x^{-1}\otimes x)= \footnote{It is enough consider the unique extention of
$\star$ to $\langle\Delta{\cal F}_I\rangle$, coherently with $\ast$ (see $(\diamondsuit),(\heartsuit)$) and 5.
of § 1.}\ x^{-1}\ast x=1=(\eta\circ\varepsilon)(x)\big(=(\mbox{id}\ \hat{\ast}\ a)(x)\big)$$for each
$x\in\Delta({\cal F}_I)$. Hence $a\ \hat{\ast}\ \mbox{id}=\mbox{id}\ \hat{\ast}\ a=\eta\circ\varepsilon$ and
therefore $a$ is an antipode, that is $(b{\cal A}_{HBJ}({\cal F}_I),a)$ is a non-commutative Hopf
$\mathbb{K}$-algebra\footnote{Furthermore, it is also a ${\ast}$-Hopf algebra with the matrix transposition,
when $\mathbb{K}=\mathbb{C}$ and card $I<\infty$.}, naturally braided by $R=1\otimes1$. We will denote it by
${\cal H}_{\mathbb{K}}({\cal F}_I)$, and will be said the\footnote{HBJ EBBH-algebra stands for \it
Heisenberg-Born-Jordan \ Ehresmann-Baer-Brandt-Hopf algebra.} \it Heisenberg-Born-Jordan EBBH-algebra \rm(or \it
HBJ EBBH-algebra\rm).\\ \rm As already said, this last structure may be regarded as the basic structure of a
quantum group that also got, a posteriori, a more physical motivation in its name. In conclusion, in the HBJ
EBBH-algebra may be recognized the eventually quantic origins of the basilar structure of quantum group,
although it has been obtained endowing the basic EBB-groupoid algebra with a trivial structure of braided Hopf
$\mathbb{K}$-algebra. There are further, well-known, (over) structures and properties on such EBB-groupoid
algebra when $I$ is finite (since we shall get a finitely generated algebra), but very few when $I$ is infinite
(that is, when the algebra is not finitely generated).\\In this first paper, we have considered the only
possible (although trivial) structure of braided non-commutative Hopf algebra on such EBB-groupoid algebra, when
$I$ is infinite (no finitely generation).\\In a further paper, we shall try to find other non-trivial structure
on it, in spite of its (interesting) no finitely (algebraic) generation.
\section*{References}{[An]} S. Antoci, \it Quando la Fisica parlava Tedesco,
\rm Quaderni del Gruppo Nazionale di Fisica Matematica (GNFM) del CNR, Firenze, 2004.\\{[As]} P. Aschieri, \it
On the Geometry of Inhomogeneous Quantum Groups, \rm Tesi di Perfezionamento, Scuola Normale Superiore (SNS),
Pisa, 1999.\\{[AM]} M.F. Atiyah, I.G. Macdonald, \it Introduction to Commutative Algebra, \rm Addison-Wesley
Publishing Comp., Reading (Massachusetts), 1969 (tr. it. \it Introduzione all'Algebra Commutativa, \rm
Feltrinelli, Milano, 1981).\\{[Ba]} A. Baer, \it Zur Einf\"{u}hrung des Scharbegriffs, \rm Journal f\"{u}r die
Reine und Angewandte Mathematik, 160 (1929) 199-207.\\{[BFFLS]} F. Bayen, M. Flato, C. Fronsdal, A.
Lichnerowicz, D. Sternheimer, \it Deformation Theory and Quantization I, II, \rm Annals of Physics, 111 (1978)
61-110, 111-151.\\{[BJ]} M. Born, P. Jordan, Zeitschrift f\"{u}r Physik, 34 (1925) 858.\\{[Bou]} N. Bourbaki,
\it Alg\`{e}bre I, \rm Hermann, Paris, 1970.\\{[Br]} W. Brandt, \it \"{U}ber eine verallgemeineurung des
gruppenbegriffes, \rm Mathematischen Annalen, 96 (1926) 360-366.\\{[CW]} A. Cannas da Silva, A. Weinstein, \it
Geometric Models for Noncommutative Algebras, \rm University of California at Berkeley, Berkeley, 1998.\\{[CCP]}
P. Caldirola, R. Cirelli, G.M. Prosperi, \it Introduzione alla Fisica Teorica, \rm UTET, Torino, 1982.\\{[CP]}
V. Chari, A. Pressley, \it A Guide to Quantum Groups, \rm Cambridge University Press, Cambridge, 1995.\\ {[Co]}
P.M. Cohn, \it Universal Algebra, \rm Harper and Row, New York, 1965 (tr. it. \it Algebra Universale, \rm
Feltrinelli, Milano, 1971).\\{[Co1]} A. Connes, \it Géométrie non commutative, \rm InterEditions, Paris,
1990.\\{[Co2]} A. Connes, \it Noncommutative Geometry, \rm in: \it Symposium on the Current State and Prospects
of Mathematics, Barcelona, June, 1991, \rm Lecture Notes in Mathematics, Springer-Verlag, Berlin, 1992.\\{[Co3]}
A. Connes, \it Noncommutative Geometry, \rm Academic Press, New York, 1994.\\{[Di1]} P.A.M. Dirac, Proceedings
of the Royal Society of London, A109 (1925) 642.\\{[Di2]} P.A.M. Dirac, \it The Principles of Quantum Mechanics,
\rm Clarendon Press, London, 1931.\\{[Dr1]} V.G. Drinfeld, \it Hopf algebras and quantum Yang-Baxter equation,
\rm Soviet Math. Dokl., 32 (1985) 254-258.\\ {[Dr2]} V.G. Drinfeld, \it Quantum groups, \rm in: \it Proceedings
of the International Congress of Mathematicians, Berkeley, 1986, \rm pp. 798-820, American Mathematical Society,
Providence, Rod Island, 1987.\\{[Ec]} C. Eckart, The Physical Review, 28 (1926) 711.\\{[Eh]} C. Ehresmann, \it
Gattungen von lokalen Strukturen, \rm Jahresbericht der Deutschen Math. Vereinigung, 60 (1957) {49-77}.\\{[EM]}
S. Eilenberg, S. MacLane, \it General theory of natural equivalence, \rm Trans. Amer. Math. Soc., 58 (1945)
231-294.\\{[ES]} S. Eilenberg, N. Steenrod, \it Foundations of Algebraic Topology, \rm Princeton University
Press, Princeton, 1952.\\{[Em]} G.G.Emch, \it Mathematical and Conceptual Foundations of 20th-Century Physics,
\rm North-Holland, Amsterdam, 1984.\\{[Fa]} L.D. Faddeev, \it Integrable models in (1+1)-dimensional QFT, \rm
Les Houches 1982, Elsevier Science Publishers, 1984.\\{[FST]} L.D. Faddeev,
E.K. Sklyanin, L.A. Takhtajan, \it The quantum inverse problem, \rm Theor. Math. Phys., 40 (1979) 194-220.\\
{[FT]} L. Faddeev, L.A. Takhtajan, \it Hamiltonian Approach to Solitons Theory, \rm Springer-Verlag, Berlin-New
York, 1987.\\{[Fe]} B. Ferretti, \it Le Radici Classiche della Meccanica Quantica, \rm Boringhieri, Torino,
1980.\\{[FH]} J. Franck, G. Hertz, Verh. D. Phys. Ges., 16 (1914) 457, 512.\\{[Gi]} E. Giordano, \it Origini e
Sviluppi della Meccanica Ondulatoria e della Meccanica delle Matrici, \rm Istituto di Fisica dell'Università di
Parma, Parma, 1974.\\{[Go]} S. Golab, \it \"{U}ber den Begriff der Pseudogrouppe von Transformationen, \rm Math.
Annalen, 116 (1939) 768-780.\\{[Gr]} K. Gruenberg, \it Una Introduzione all'Algebra Omologica, \rm Pitagora
Editrice, Bologna, 2002.\\{[HBJ]} W. Heisenberg, M. Born, P. Jordan, Zeitschrift f\"{u}r Physik, 35 (1926)
557.\\{[He1]} W. Heisenberg, Zeitschrift f\"{u}r Physik, 33 (1925) 879;\\{[He2]} W. Heisenberg, \it Die
Physikalischen Prinzipien der Quantentheorie, \rm S. Hirzel Verlag, Lipsia, 1930 (tr. it. \it I Principi Fisici
della Teoria dei Quanti, \rm Boringhieri, Torino, 1976).\\{[Her]} G. Herzberg, \it Atomspektren und
Atomstruktur, \rm Th. Steinkopff, Dresda, 1936 (tr. it. \it Spettri Atomici e Struttura Atomica, \rm
Boringhieri, Torino, 1961).\\{[Hi]} P.J. Higgins, \it Categories and Groupoids, \rm D. Van Nostrand Publ. Comp.,
Amsterdam, 1971.\\{[Hu]} F. Hund, \it Geschichte der Quantentheorie, \rm Bibliographishes Institut AG, Zurigo,
1975 (tr. it. \it Storia della Teoria dei Quanti, \rm Boringhieri, Torino, 1980).\\{[Ji1]} M. Jimbo, \it A
q-difference analogue of $U(\mathfrak{g})$ and the Yang-Baxter equation, \rm Lett. Math. Phys., 10 (1985)
63-69.\\{[Ji2]} M. Jimbo, \it A q-analogue of $U(\mathfrak{gl}(N+1))$, Hecke algebra and the Yang-Baxter
equation, \rm Lett. Math. Phys., 11 (1986) 247-252.\\{[Ka]} C. Kassel, \it Quantum Groups, \rm Springer-Verlag,
Berlin, 1995.\\{[KS]} A. Klimyk, K. Schmudgen, \it Quantum Groups and Their Representations, \rm
Springer-Verlag, Berlin, 1997.\\{[KR]} P.P. Kulish, N.Y. Reshetikhin, \it The quantum linear problem for the
sine-Gordon equation and highest representations, \rm J. Soviet Math., 23 (1983) 2435-2441.\\{[KuS]} P.P.
Kulish, E.K. Sklyanin, \it Solutions of the Yang-Baxter equation, \rm J. Soviet Math., 19 (1982)
1596-1620.\\{[La]} N.P. Landsman, \it Mathematical Topics Between Classical and Quantum Mechanics, \rm Spinger,
New York 1998.\\{[Lo]} P.O. L\"{o}wdin, \it Group Algebra, Convolution Algebra, and applications to Quantum
Mechanics, \rm Reviews of Modern Physics, 39 (2) (1967) 259-287.\\{[Lu]} G. Ludwig, \it Wave Mechanics, \rm
Pergamon Press, London, 1968.\\{[Mac]} C.C. Macduffee, \it The Theory of Matrices, \rm Chelsea Publ. Company,
New York, 1946.\\{[Mak]} K. Mackenzie, \it Lie groupoids and Lie algebroids in Differential Geometry, \rm
Cambridge University Press, Cambridge, 1987.\\{[MacL]} S. MacLane, \it Categories for the Working Mathematician,
\rm Springer-Verlag, New York, 1971.\\{[Ma]} I.J. Maddox, \it Elements of Functional Analysis, \rm Cambridge
University Press, Cambridge, 1970.\\{[Md]} S. Majid, \it Foundations of Quantum Group Theory, \rm Cambridge
University Press, Cambridge, 1995.\\{[Mn]} Y.I. Manin, \it Quantum Groups and Non-Commutative Geometry, \rm
Publications du CRM de l'Universite de Montreal, Montreal, 1988.\\{[Mo]} W.J. Moore, \it Physical Chemistry, \rm
tr. it. \it Chimica Fisica, \rm Piccin Editore, Padova, 1980.\\{[Ni]} A. Nijenhuis, \it Theory of the geometric
object, \rm Amsterdam, 1952.\\{[No]} D.G. Northcott, \it Ideal Theory, \rm Cambridge University Press,
Cambridge, 1953.\\{[Pi]} B. Pini, \it Primo Corso di Algebra, \rm Editrice CLUEB, Bologna, 1967.\\{[Sch]} E.
Schr\"{o}dinger, Annalen der Physik, 79 (1926) 361, 489, 734; 80 (1926) 437; 81 (1926) 109.\\{[Sh]} I. Schur,
Sitzungsberichte der Preuss. Akademie der Wissenschaften, Berlin (1905) 406-432.\\{[St]} F. Strocchi, \it An
Introduction to the Mathematical Structure of Quantum Mechanics, \rm lecture notes given at the Scuola Normale
Superiore, Pisa, 1996, or: F. Strocchi, \it An Introduction to the Mathematical Structure of Quantum Mechanics,
\rm World Scientific Publishing Company, Singapore, 2005.\\{[Vr]} A. Vretblad, \it Fourier Analysis and its
Applications, \rm GTM 223, Springer, Berlin, 2003.\\{[Wa]} W. Waliszewki, \it Categories, groupoids,
pseudogroups and analytic structures, \rm Rozprawy Matematyczne, Instytut Matematyczny Polskiej Akademii Nauk,
XLV, Warszawa, 1965.\\{[We]} H. Weyl, \it The Classical Groups, \rm Princeton University Press, Princeton,
1946.\\{[Wo]} S.L. Woronowicz, \it Compact matrix pseudo-groups, \rm Comm. Math. Phys., 111 (1987) 613-665.

\end{document}